\documentclass[a4paper,11pt]{article}
\pdfoutput=1 % if your are submitting a pdflatex (i.e. if you have
\usepackage{jheppub} % for details on the use of the package, please
\usepackage[T1]{fontenc} % if needed

\usepackage{comment}
\usepackage{graphicx,rotating,amsmath,amsfonts,amssymb}

\newcommand{\be}{\begin{equation}}
\newcommand{\ee}{\end{equation}}
\newcommand{\br}{\begin{eqnarray}}
\newcommand{\bea}{\begin{eqnarray}}
\newcommand{\eea}{\end{eqnarray}}
\newcommand{\er}{\end{eqnarray}}
\newcommand{\ba}{\begin{array}}
\newcommand{\ea}{\end{array}}
\newcommand{\nn}{\nonumber}
\newcommand{\bi}{\begin{itemize}}
\newcommand{\ei}{\end{itemize}}
\newcommand{\bn}{\begin{enumerate}}
\newcommand{\en}{\end{enumerate}}
\newcommand{\bc}{\begin{center}}
\newcommand{\ec}{\end{center}}

\newcommand{\Eq}[1]{Eq.~(\ref{#1})}

\newcommand{\PRexp}{2.14 \pm 0.05}

\newcommand{\beq}{\begin{equation}}
\newcommand{\eeq}{\end{equation}}

\newcommand{\gsim}{\lower.7ex\hbox{$\;\stackrel{\textstyle>}{\sim}\;$}}
\newcommand{\lsim}{\lower.7ex\hbox{$\;\stackrel{\textstyle<}{\sim}\;$}}

\title{Linear inflation from quartic potential}

\author[a]{Kristjan Kannike}

\author[a]{Antonio Racioppi}

\author[a,b]{Martti Raidal}
\affiliation[a]{National Institute of Chemical Physics and Biophysics, R\"{a}vala 10, 10143 Tallinn, Estonia}
\affiliation[b]{Institute of Physics, University of Tartu, Estonia}

\date{\today}

\emailAdd{antonio.racioppi@kbfi.ee}
\emailAdd{kristjan.kannike@cern.ch}
\emailAdd{martti.raidal@cern.ch}

\abstract{
We show that if the inflaton has a non-minimal coupling to gravity and the Planck scale is dynamically generated, the results of Coleman-Weinberg inflation are confined in between two attractor solutions: quadratic inflation, which is ruled out by the recent measurements, and linear inflation which, instead, is in the experimental allowed region.
The minimal scenario has only one free parameter -- the inflaton's non-minimal
coupling to gravity -- that determines all physical parameters such as  the tensor-to-scalar ratio and the reheating temperature of the Universe.
Should the more precise future measurements of inflationary parameters point towards linear inflation,
further interest in scale-invariant scenarios would be motivated.
}

\begin{document}

\maketitle

\section{Introduction}
After the Background Imaging of Cosmic Extragalactic Polarization (BICEP2) claimed the detection of tensor modes from large angle Cosmic Microwave Background (CMB) B-mode polarisation~\cite{Ade:2014xna}, a lot of interest was given to inflationary models with a relatively big scalar-to-tensor ratio $r$. However, the later analysis performed by the BICEP2-Planck joint collaboration showed that most of the signal measured by BICEP2~\cite{Ade:2014xna} was actually due to the dust background~\cite{Ade:2015tva}. Moreover, the updated constraints on inflation taking into account the Planck Mission polarisation measurements are restricting the allowed inflation parameter space even more, reducing the number of phenomenologically allowed models~\cite{Ade:2015lrj}. All the monomial inflaton potentials with integer powers are disfavoured, except for the one linear in the inflaton field.

However, it is not straightforward to generate a linear inflationary potential in a simple quantum field theory (QFT) framework that follows from a well-motivated
ultraviolet (UV)-completed theory. The popular QFT linear inflaton potentials appearing in the literature can be considered only as local effective descriptions of
more complicated scalar potentials~\cite{Martin:2013tda}. For example, linear inflation may appear as a limiting case of the quartic hill-top inflation~\cite{Boubekeur:2005zm}.
Since the latter potential is not bounded from below, it cannot represent a fundamental theory.
Alternatively, the linear potentials appear from the axion monodromy of wrapped branes in string compactifications~\cite{McAllister:2008hb}.
Understanding the UV-complete origin of such a potential requires the understanding of realistic string theory.

In a recent paper \cite{Rinaldi:2015yoa} it has been shown that the linear inflaton potential appears as an attractor solution in the Coleman-Weinberg~\cite{Coleman:1973jx}
inflation provided the inflaton has a non-minimal coupling to gravity.
Thus the predictions of the linear inflation for the spectral index $n_s$ and for $r$ can be achieved in the context of well-defined QFT framework from the quartic scalar potentials
without introducing any complicated {\it ad hoc} interaction or unbounded scalar potential by hand.
The aim of this work is to extend the previous discussion of \cite{Rinaldi:2015yoa}, presenting a more detailed study of the parameter space, including also a discussion about reheating. Should the more precise future measurements~\cite{Abazajian:2013vfg}
of $(n_s,\,r)$ clearly favour the linear
inflation and distinguish it from the other well-known attractor solutions such as the quadratic inflation~\cite{Linde:1983gd} or the Starobinsky inflation~\cite{Starobinsky:1980te},
further interest in scale-invariant scenarios would be motivated.
%this would strongly support the scale-invariant origin of inflation.

%This scenario is supported by experimental data from other experiments.
This scenario is inspired by recent collider data.
The discovery of the Higgs boson~\cite{Chatrchyan:2012xdj,Aad:2012tfa}, and the absence (so far) of any other new physics at the LHC experiments, have challenged the standard paradigm of naturalness of the electroweak scale. Various authors have recently proposed modified approaches to naturalness  trying to develop a theoretical framework able of explaining the co-existence and the origin of the largely separated mass scales observed in nature.
Most attempts involve, in some way or another, the idea of classical scale invariance of the standard model (SM)~\cite{Bardeen:1995kv} and the loop-level dimensional
transmutation~\cite{Coleman:1973jx}.

The concept of Coleman-Weinberg inflation is old. Already the very first papers on inflation \cite{Linde:1981mu,Albrecht:1982wi,Linde:1982zj,Ellis:1982ws,Ellis:1982dg}
considered  potentials whose shapes were dynamically induced at one loop level via the Coleman-Weinberg mechanism~\cite{Coleman:1973jx}.
 Since then  the Coleman-Weinberg inflation has been extensively studied in the context of grand unified theories
\cite{Langbein:1993ym,GonzalezDiaz:1986bu,Yokoyama:1998rw,Rehman:2008qs} and in the $U(1)_{B-L}$ extension of the
SM~\cite{Barenboim:2013wra,Okada:2013vxa}.  In those models the dynamics leading to dimensional transmutation is most commonly
assumed to be new gauge interactions beyond the SM.
However, the dimensional transmutation  can occur just due to running of some scalar quartic coupling $\lambda(\mu)\phi^4$
due to the $\phi$ coupling to some other scalar field~\cite{Hempfling:1996ht,Gabrielli:2013hma},
generating a non-trivial inflaton potential as demonstrated in Ref.~\cite{Kannike:2014mia}.
In this paper we study this type of inflation models\footnote{For a different use of the Coleman-Weinberg mechanism and its connection to Composite Higgs models, see \cite {Croon:2015fza,Croon:2015naa}. } in the presence of a non-minimal coupling to gravity.

The Coleman-Weinberg inflation in the presence of a non-minimal coupling to gravity has previously been studied by several authors ~\cite{Panotopoulos:2014hwa,Okada:2014lxa}.
In those models the Planck scale is explicitly written in the Lagrangian.
In our case, instead, we will assume that also the Planck scale is dynamically generated via the inflaton vacuum expectation value (VEV) and its non-minimal coupling to gravity,
following closely the ideology in~\cite{Kannike:2015apa}. This will imply a completely different phenomenology and lead to the results stated above.

Quantum effects for the quartic $\lambda(\mu)\phi^4$ inflation with non-minimal coupling to gravity are known up to two loops~\cite{Inagaki:2015fva}.
In this work the Planck scale is also an explicit parameter, {\it i.e.,} the gravity is explicitly not scale invariant, and the loop corrections to the inflaton potential
are dominated by the inflaton self-interactions.
In our case the dominant loop effect is induced by the inflaton portal-type interaction with the dark sector. This will imply, again, different phenomenological results.
%Therefore, to our best knowledge, the results presented here are new!

%\bigskip

The organisation of the paper is the following.
In section~\ref{sec:CSI_and_inflation} we present general discussion of  the Coleman-Weinberg inflation with the inflaton non-minimal coupling to gravity. In section \ref{sec:mod:ind} we treat the inflationary dynamics in a model independent way assuming a generic  potential arising from the non-minimal Coleman-Weinberg inflation.
In section \ref{sec:mod:dip} we present the minimal model that generates the  non-minimal Coleman-Weinberg inflation and
present results for two limiting cases.
In section~\ref{sec:Reheating} we study reheating of the Universe in our scenario.
%and discuss its implications for leptogenesis.
We conclude in section~\ref{sec:Discussion}. Technical details of our computations are presented in Appendix~\ref{appendix}.

%%%%%%%%%%%%%%%%%%%%%%%
\section{Coleman-Weinberg potential and non-minimal coupling to gravity} \label{sec:CSI_and_inflation}
%%%%%%%%%%%%%%%%%%

In \cite{Kannike:2014mia} we presented a simple model of two scalar fields, $\phi$ and $\sigma,$ and
showed that at 1-loop level this model generates the Coleman-Weinberg effective inflation potential of the form
\be
  V_{\rm CW} =\Lambda^4 +\frac{1}{8} \beta_{\lambda_\phi} \left( \ln \frac{\phi^{2}}{v_\phi^{2}} - \frac{1}{2} \right) \phi^4 , \label{eq:Veff:Jordan}
\ee
where
\be
 \beta_{\lambda_\phi} \simeq \frac{\lambda_{\phi \sigma}^2}{32\pi^2},
\ee
is the beta function of the inflaton quartic coupling $\lambda_\phi$, $v_\phi$ is the inflaton vacuum expectation value (VEV), and $\lambda_{\phi \sigma}$ the portal coupling between the inflation $\phi$ and the field $\sigma$ (the latter can be considered as a non-varying field since its running is strongly suppressed \cite{Kannike:2014mia}).
In \Eq{eq:Veff:Jordan} the $\Lambda$ term is a constant potential added in order to realise $V(v_\phi)=0$ in order to solve the cosmological constant problem and
 avoid problems related to eternal inflation \cite{Guth:2007ng}.
  While particle physics observables depend only on the difference of the potential, gravity couples to the absolute scale, creating the cosmological constant problem that so far does not have a commonly accepted elegant solution. In the following we view the existence of $\Lambda$ as a phenomenological necessity and accept the fine tuning associated with it.

The potential in \Eq{eq:Veff:Jordan} has an attractor solution, the quadratic potential $m^2\phi^2,$ that is now disfavoured by
 the new Planck and BICEP2 data \cite{Ade:2015lrj}. However, for a generic parameter space this model can still be compatible with data within $2\sigma.$
The whole construction above is based on the assumption of the classical scale invariance as a property of matter, but not gravity, introducing an explicit Planck mass $M_P$
as a constant. The inflaton was minimally coupled to gravity. To extend the argument of classical scale invariance,  in \cite{Kannike:2015apa} we considered adimensionality as a property of the full classical action, forbidding any tree-level $M_P$ and generating dynamically the Planck scale via non-minimal coupling $\xi\phi^2 R$.
In this work we consider the same scenario where the Planck scale is completely generated via the non-minimal coupling of the inflaton with the scalar curvature.
As the only known solution to the cosmological constant problem is fine tuning, we also assume a tree-level constant $\Lambda$ without specifying its origin.

Given the assumptions, the Lagrangian for such a scenario in the Jordan frame is:
\begin{equation}
  \sqrt{- g^{J}} \mathcal{L}^{J} = \sqrt{- g^{J}} \left[ \mathcal{L}_{\rm SM} - \frac{\xi_\phi}{2} \phi^{2} R
  + \frac{(\partial \phi)^{2}}{2} - V^J_{\rm eff}(\phi) + \mathcal{L}^{J}(\sigma, \psi, A_\mu) + \Lambda^4\right] ,
   \label{eq:Jordan:Lagrangian}
\end{equation}
where $\mathcal{L}_{\rm SM} $ is the SM Lagrangian, $V^J_{\rm eff}(\phi)$ is a generic 1-loop scalar potential generated from a classical scale invariant potential,\footnote{In presence of an Einstein-Hilbert term $-\frac{M^2}{2} R$, the Lagrangian (\ref{eq:Jordan:Lagrangian}) can be considered as a limiting case $M \ll M_P$, which can be easily realised with the \emph{natural} choice $M \sim \Lambda$ (cf. Fig. \ref{fig:TRHvsxi}).}
 \begin{equation}
  V^J_{\rm eff}(\phi) = \frac{1}{4} \lambda_\phi (\phi) \phi^4 \label{eq:Veff:J},
 \end{equation}
 and $\mathcal{L}^{J}(\sigma, \psi, A_\mu)$ are the Lagrangian terms involving the extra scalar, fermion or vector matter fields that we do not specify at the moment.
 The potential $V^J_{\rm eff}(\phi)$ is already projected in the direction where inflation happens,  {\it i.e.,} where other scalar field values are frozen to their minimum position.
 A concrete example will be given later.

 The Lagrangian (\ref{eq:Jordan:Lagrangian}) lacks an Einstein-Hilbert term. This has to be generated by inducing a different from zero VEV to the inflaton. Therefore, to generate the Planck scale, the VEV $v_\phi$ of the inflaton field must be given by
\begin{equation}
  v_{\phi}^{2} = \frac{M_P^{2}}{\xi_\phi}.
  \label{eq:v:phi:Planck:mass}
\end{equation}
Such a relation is the main motivation why our construction does not reproduce the results of the non-minimally coupled to gravity Higgs inflation model \cite{Bezrukov:2007ep}, i.e. ranging from quartic to Starobinsky inflation according to the value of the non-minimal coupling to gravity. They include an the explicit Einstein-Hilbert term $-\frac{M_P^2}{2} R$ in the Lagrangian, while in our model it is dynamically generated via the inflaton VEV.
Note that such a relation automatically implies that $\xi_\phi$ can only take positive values, forbidding the conformal value $\xi_\phi=-1/6$.
The general equation for minimising the potential (\ref{eq:Jordan:Lagrangian}), (\ref{eq:Veff:J}) at $v_\phi \neq 0$ is
\begin{equation}
 \frac{1}{4} \beta_{\lambda_\phi} (v_\phi) + \lambda_\phi (v_\phi) =0.
\end{equation}
Therefore, several possibilities are open according to how we solve the equation. The case $\beta_{\lambda_\phi} (v_\phi) =\lambda_\phi (v_\phi)=0$ was already studied in \cite{Kannike:2015apa}, implying $\Lambda=0$. Thus we remain with two additional choices:
\begin{eqnarray}
\text{i)} & \beta_{\lambda_\phi} (v_\phi) < 0, \ \lambda_\phi (v_\phi)>0 , \\
\text{ii)} & \beta_{\lambda_\phi} (v_\phi) > 0, \ \lambda_\phi (v_\phi)<0 \label{eq:RGE:bound:cond} .
\end{eqnarray}
Since $v_\phi$ has to be the absolute minimum of the potential, the only phenomenologically allowed choice is (\ref{eq:RGE:bound:cond}). Hence the need of the constant $\Lambda$, so that $ V^J_\phi(v_\phi)=0$.  The choice (\ref{eq:RGE:bound:cond}) has also consequences involving the matter content of the model. In order to have a stable $V_\phi^J$, $\lambda_\phi$ must flip sign at a certain scale $\mu_0 > v_\phi$ and remain positive. However this is not achievable only considering the inflaton loop-self corrections. Given the boundary condition $\lambda_\phi (v_\phi)<0$, the self-corrections of $\lambda_\phi$ will push the quartic coupling itself to bigger but negative values. Therefore, the stabilisation of the scalar potential combined with (\ref{eq:RGE:bound:cond}) requires the presence of some new particle.

%%%%%%%%%%%%%%%%%%%%
\section{General results for non-minimal Coleman-Weinberg inflation} \label{sec:mod:ind}
%%%%%%%%%%%%%%%%%%%%%

\begin{comment}
In this case the running of the portal $\lambda_{\phi \sigma}$ is quite suppressed so that we can safely approximate
\begin{equation}
 \lambda_{\phi \sigma} (\mu) \simeq  \lambda _{\phi\sigma } (v_{\phi }).
\end{equation}
Since $\lambda_{\phi \sigma}$ is essentially constant, we drop its $\mu$ dependence for the rest of this subsection.
\end{comment}

We start with studying general properties of the scenario outlined in the previous section.
For that purpose we do not need to specify full model details. To the contrary, we show that
our results are rather model independent and applicable to a wide class of non-minimal Coleman-Weinberg  inflaton potentials
for which the following holds:
\begin{itemize}
\item the beta function of the inflaton quartic coupling is slowly varying and its running can be neglected, implying
\begin{equation}
 \lambda_\phi (\mu) \simeq \beta_{\lambda_\phi} \ln \frac{\mu}{\mu^*} , \label{eq:lambda:phi:approx}
\end{equation}
where $\mu^*$ is the scale at which the quartic coupling is zero;
\item the non-minimal coupling $\xi_\phi$ is slowly varying, implying $\beta_{\xi_\phi}\simeq 0$.
\end{itemize}
Such a limiting solution is shared by a wide class of physically interesting  Coleman-Weinberg inflation models, independently of the exact matter content of $\mathcal{L}^{J}(\sigma, \psi, A_\mu)$, that is scalars, fermions or vectors. As far as the two conditions given before hold, our results for the behaviour of $r$,
$n_s$ and the inflaton mass, will be quite general and independent on the actual Coleman-Weinberg realisation.
We present a most minimal explicit model possessing those properties in the next section.

Given the assumptions, it can be easily checked that the 1-loop scalar potential in the Jordan frame is given by eq. (\ref{eq:Veff:Jordan}). It is convenient to rewrite it in a more convenient form. As we said before,  the constant $\Lambda$ should be fixed in order to achieve $V_{\rm eff} (v_\phi)=0$.
Solving such a constraint as a function of $\beta_{\lambda_\phi} $ we get
\begin{equation}
 \beta_{\lambda_\phi}  = 16 \frac{\Lambda^4}{v_\phi^4} ,
\label{eq:portal}
\end{equation}
and the effective potential can be rewritten as
\begin{equation}
  V_{\rm eff} =
\Lambda ^4 \left\{ 1 + \left[ 2 \ln \left(\frac{\phi^2}{v_{\phi }^2}\right) -1 \right] \frac{\phi^4}{v_{\phi }^4} \right\} \label{eq:Veff:Jordan:Lambda}.
\end{equation}
It is well known that inflationary observables can be equivalently computed (see \cite{Chiba:2008ia,Chiba:2008rp} and references therein) in the Jordan frame possessing  the non-minimal coupling with gravity, or in the Einstein frame possessing the canonical Einstein-Hilbert action of gravity. It is just more convenient to move to the Einstein frame, since the interpretation of the results is more intuitive and simple. The change of frame is achieved with the Weyl transformation
\begin{equation}
  g_{\mu\nu}^E = \Omega(\phi)^{2} g_{\mu\nu}, \qquad \hbox{where}\qquad\Omega(\phi)^{2} = \frac{\phi^{2}}{v_\phi^{2}}.
\label{eq:conformal}
\end{equation}
The Einstein frame scalar potential is then given by
\begin{equation}
  V_E(\phi) = \frac{V_{\rm eff}(\phi)}{\Omega(\phi)^{4}}.
  \label{eq:VE:J}
\end{equation}
As a last step we must to rewrite $V_E$ in terms of the canonically normalised field $\phi_{E}$ in the Einstein frame,
\begin{equation}
\phi_E = \sqrt{\frac{1+6 \xi_\phi}{\xi_\phi }} M_P \ln \frac{\phi}{v_\phi},
\label{eq:phiE}
\end{equation}
or equivalently
\begin{equation}
  \phi = v_{\phi}  e^{\sqrt{\frac{\xi_\phi}{1+6 \xi_\phi}} \frac{\phi_E}{M_P}} .
  \label{eq:phiJ}
\end{equation}
Therefore combining eqs. (\ref{eq:VE:J}), (\ref{eq:Veff:Jordan:Lambda}), (\ref{eq:phiJ}) and (\ref{eq:v:phi:Planck:mass}) we get
\begin{equation}
V_E(\phi_E)=\Lambda ^4 \left(4\sqrt{\frac{\xi_\phi}{1+6 \xi_\phi}} \frac{\phi_E}{M_P}+e^{-4\sqrt{\frac{\xi_\phi}{1+6 \xi_\phi}} \frac{\phi_E}{M_P}}-1\right) .  \label{eq:Veff:Einstein}
\end{equation}

\begin{figure}[t!]
\centering
 \includegraphics[width=0.7\textwidth]{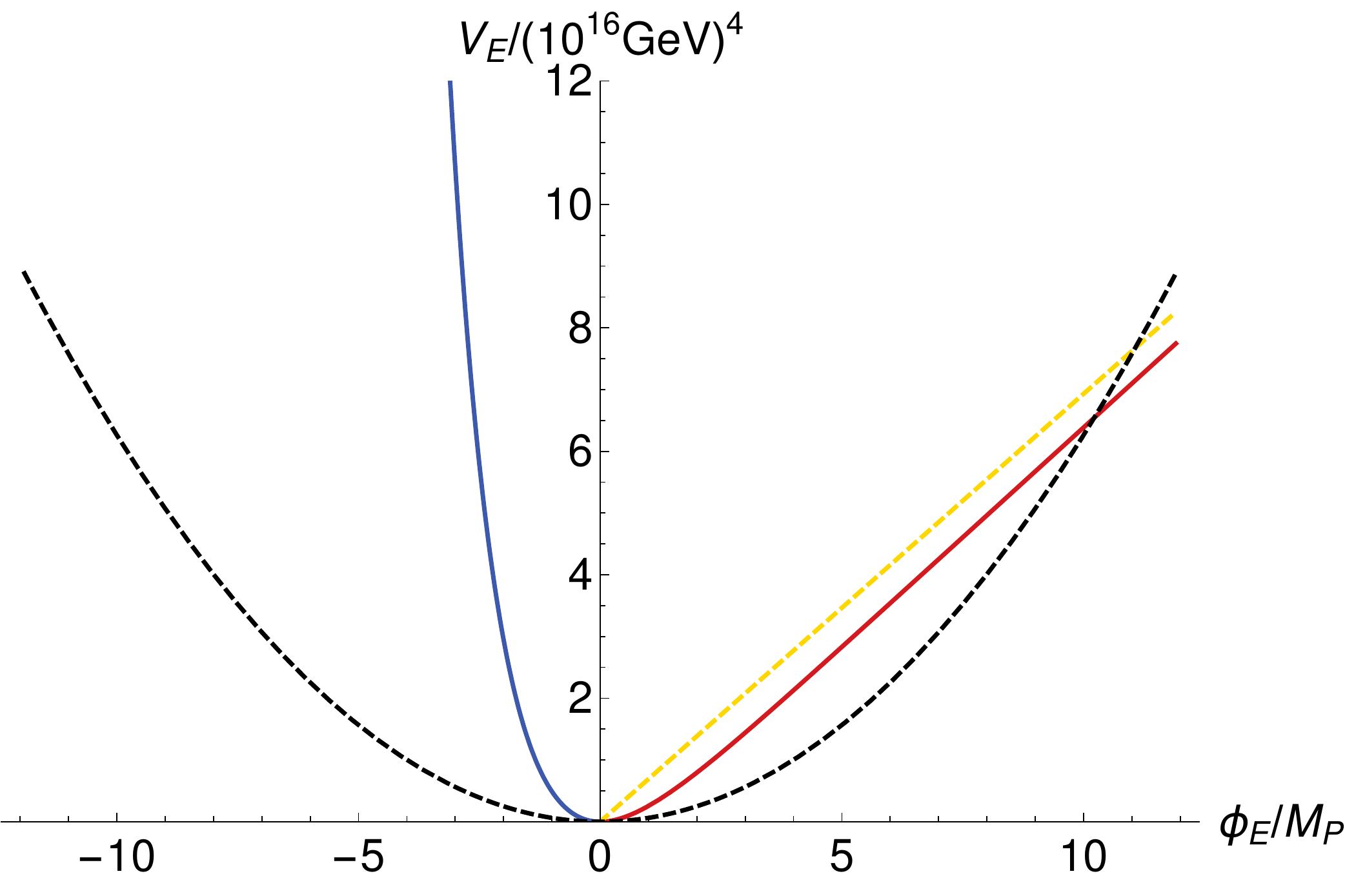}
 \caption{Inflaton potential $V_E(\phi_E)$ as a function of the field value for fixed $\Lambda \simeq 9.2 \times 10^{15}$ GeV and $\xi_\phi=0.1$.  The blue (red) full line corresponds to
 $\phi_E <0 (>0)$ where the exponential (linear) term dominates. The black dashed line depicts the quadratic limit and the yellow dashed the linear limit.}
 \label{Vfig}
\end{figure}

To exemplify our results, we plot in Fig. \ref{Vfig} the potential $V_E(\phi_E)$ for fixed values of the free parameters $\Lambda \simeq 9.2 \times 10^{15}$ GeV and $\xi_\phi=0.1$.
The shape of the potential is quite interesting and allows for two simple limiting cases. The first one is the limit $\xi_\phi\to 0$ which gives immediately
\begin{equation}
V_E(\phi_E) \Big|_{\xi_\phi \to 0} \sim 8 \xi_\phi \frac{\Lambda ^4}{M_P^2}  \phi _E^2 ,\label{eq:Veff:Einstein:quad}
\end{equation}
which is nothing but a quadratic potential that is phenomenologically disfavoured in the light of new Planck data \cite{Ade:2015lrj}. The other limiting case is $\xi_\phi\to +\infty$, which gives two different predictions  for different field values $\phi_E>0$ or $\phi_E<0$:
\begin{equation}
V_E(\phi_E) \Big|_{\xi_\phi \to +\infty} \sim
\left\{
\begin{array}{c}
\Lambda ^4 e^{\sqrt{\frac{8}{3}} \frac{|\phi_E|}{M_P}} \quad \phi_E<0, \\
\Lambda ^4 \sqrt{\frac{8}{3}} \frac{\phi_E}{M_P} \quad \phi_E>0,
\end{array}
\right.  \label{eq:Veff:Einstein:big:xi}
\end{equation}
where we used the fact that inflation happens for trans-Planckian field values $|\phi_E| \gg M_P$. For $\phi_E<0$ (small field value in the Jordan frame i.e. $\phi < v_\phi$) we see that the exponential term dominates, being already ruled out phenomenologically, while for $\phi_E>0$ (large field value in the Jordan frame i.e. $\phi > v_\phi$) the linear term dominates. The latter is in good accordance with Planck and BICEP2 data.
The existence of linear limit in the Coleman-Weinberg inflation is the main result of this paper.

The computation of the slow-roll parameters and the number of $e$-folds can be performed exactly using the full potential eq.~(\ref{eq:Veff:Einstein}). We give more details about it in the Appendix \ref{appendix}. In Fig. \ref{Fig:rvsn} we present the results for the tensor-to-scalar ratio $r$ for $N \in [50,60]$ $e$-folds as a function of $n_s$ (left panel) and as a function of $\xi_\phi$ (right panel).  The blue region represents the negative field inflation while the red one the positive field inflation. For reference we also plot predictions of quadratic (black) and linear (yellow) potentials.
The light green areas present the 1,2$\sigma$ best fits to the { Planck, BICEP2/Keck} data~\cite{Ade:2015tva,Ade:2015fwj,Ade:2015xua,Ade:2015lrj}.

\begin{figure}[t]
\centering
 \includegraphics[width=0.45\textwidth]{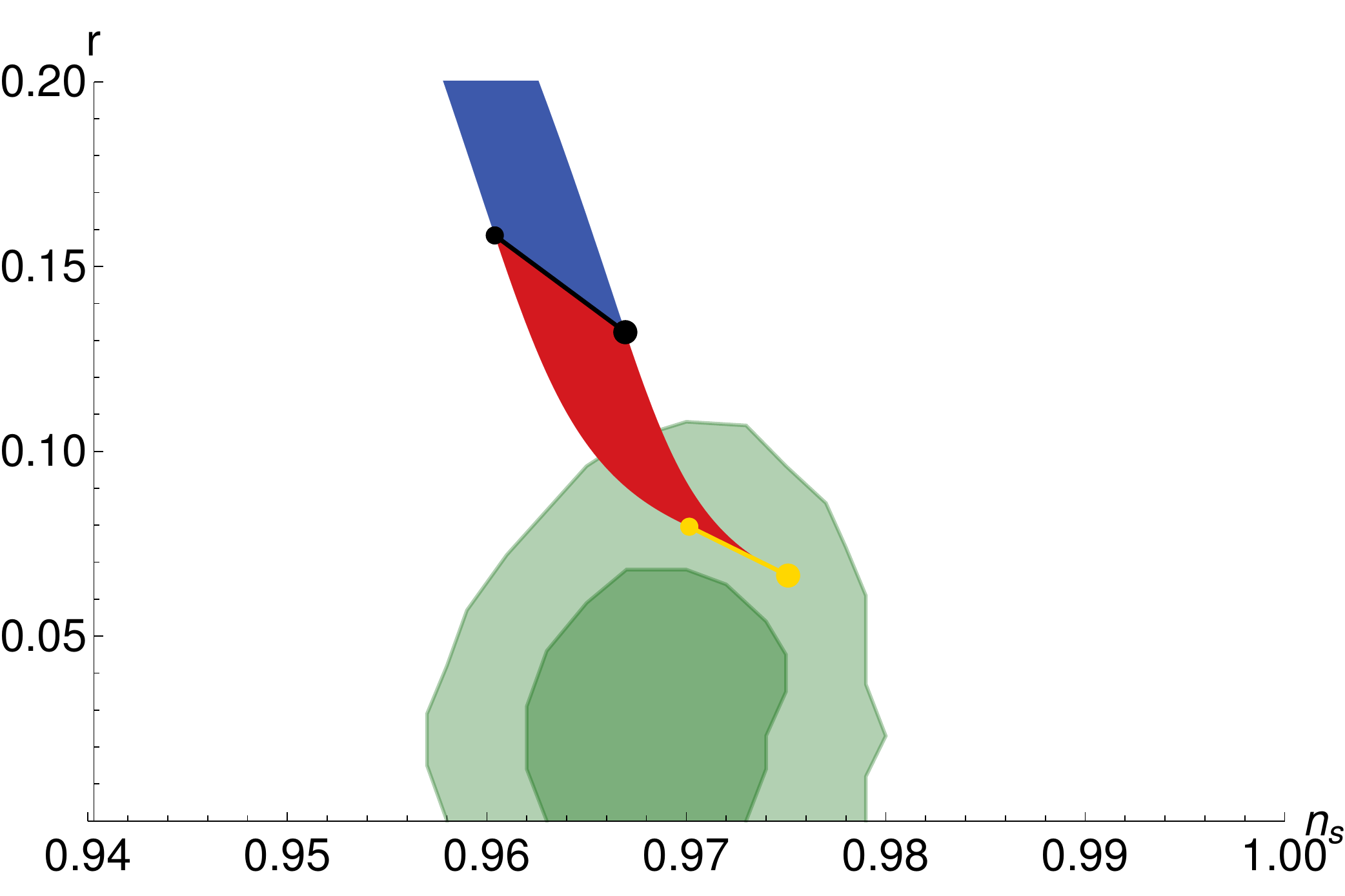}
 \quad
 \includegraphics[width=0.45\textwidth]{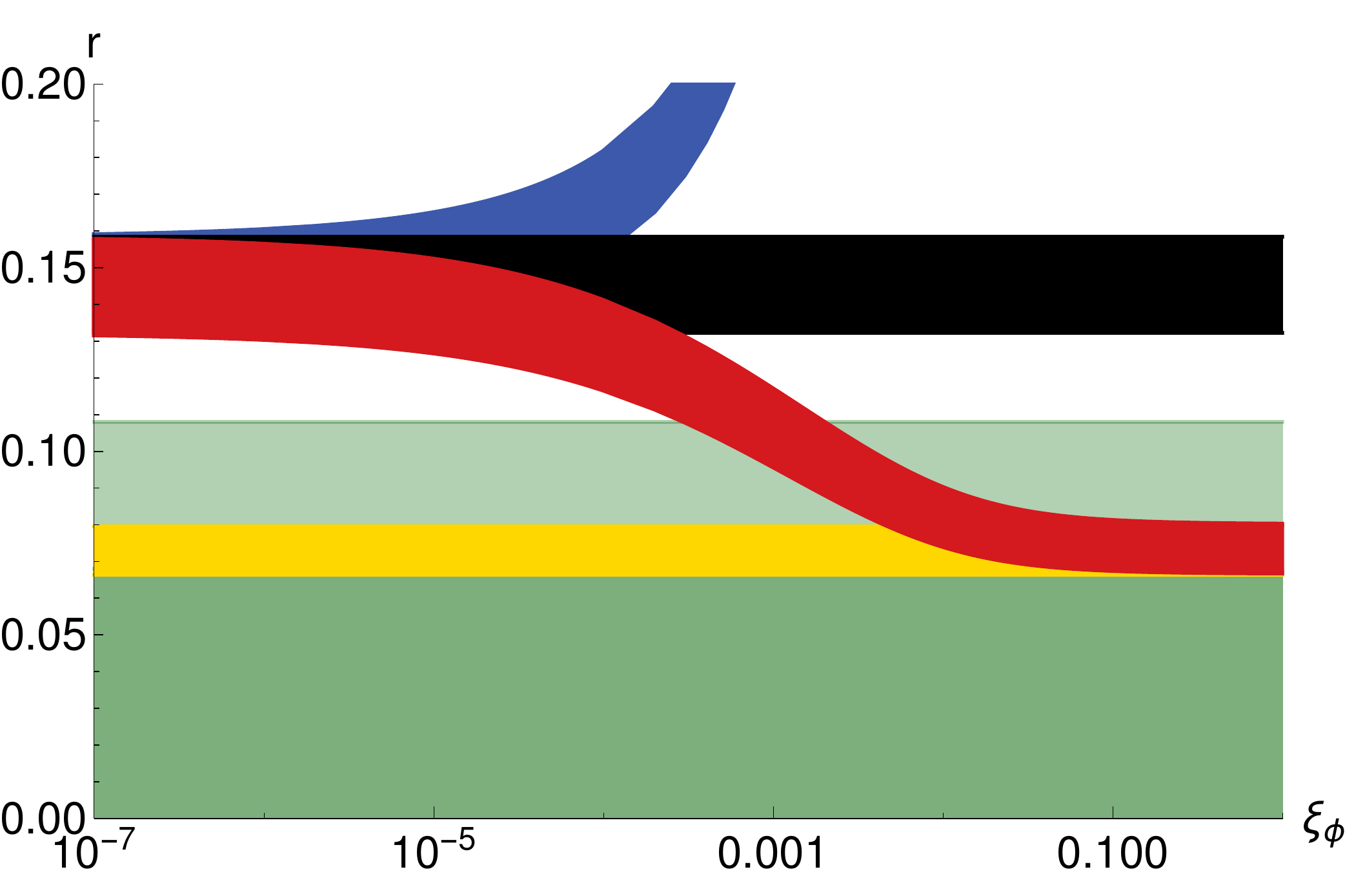}
 \caption{Predictions for the tensor-to-scalar ratio $r$ for $N \in [50,60]$ $e$-folds as a function of $n_s$ (left panel) and as a function of $\xi_\phi$ (right panel).  The blue region represents the negative field inflation while the red one the positive field inflation.  For reference we also plot predictions of  $m^2 \phi_E^2$ (black) and $m^3 \phi_E$ (yellow) potentials. The light green areas present the 1,2$\sigma$ best fits to the { Planck and BICEP2/Keck} data~\cite{Ade:2015tva,Ade:2015fwj,Ade:2015xua,Ade:2015lrj}.
}
 \label{Fig:rvsn}
\end{figure}

This potential has formally only two parameters: $\Lambda$, that should be fixed by normalisation, and $\xi_\phi$ that is the only free dynamical parameter. Since the slow-roll parameters, in particular the observables $r$ and $n_s,$ are normalisation independent, the results of Fig.~\ref{Fig:rvsn} depend only on the value of $\xi_\phi$.
For very small values of $\xi_\phi$ ($\lesssim 10^{-6}$) the predictions of this model essentially coincide with the ones of quadratic inflation.
Increasing the value of $\xi_\phi$, $r$ departs from the quadratic limit, starting to saturate the linear limit\footnote{A similar interpolation between quadratic and linear inflation was also obtained in \cite{Escobar:2015fda}, with a much more complicated model involving supersymmetry, $D$-branes and compactification of extra dimensions.} for $\xi_\phi\gtrsim 0.1$ in case of positive-field inflation. We see that the values of $r$ approach the Planck favoured region for $\xi_\phi \gtrsim 0.001$. It is interesting to notice that the linear limit is reached
for moderately small values of  $\xi_\phi$. Therefore, we used an upper bound $\xi_\phi\lsim 1$ in our examples.

\begin{figure}[t]
\centering
 \includegraphics[width=0.45\textwidth]{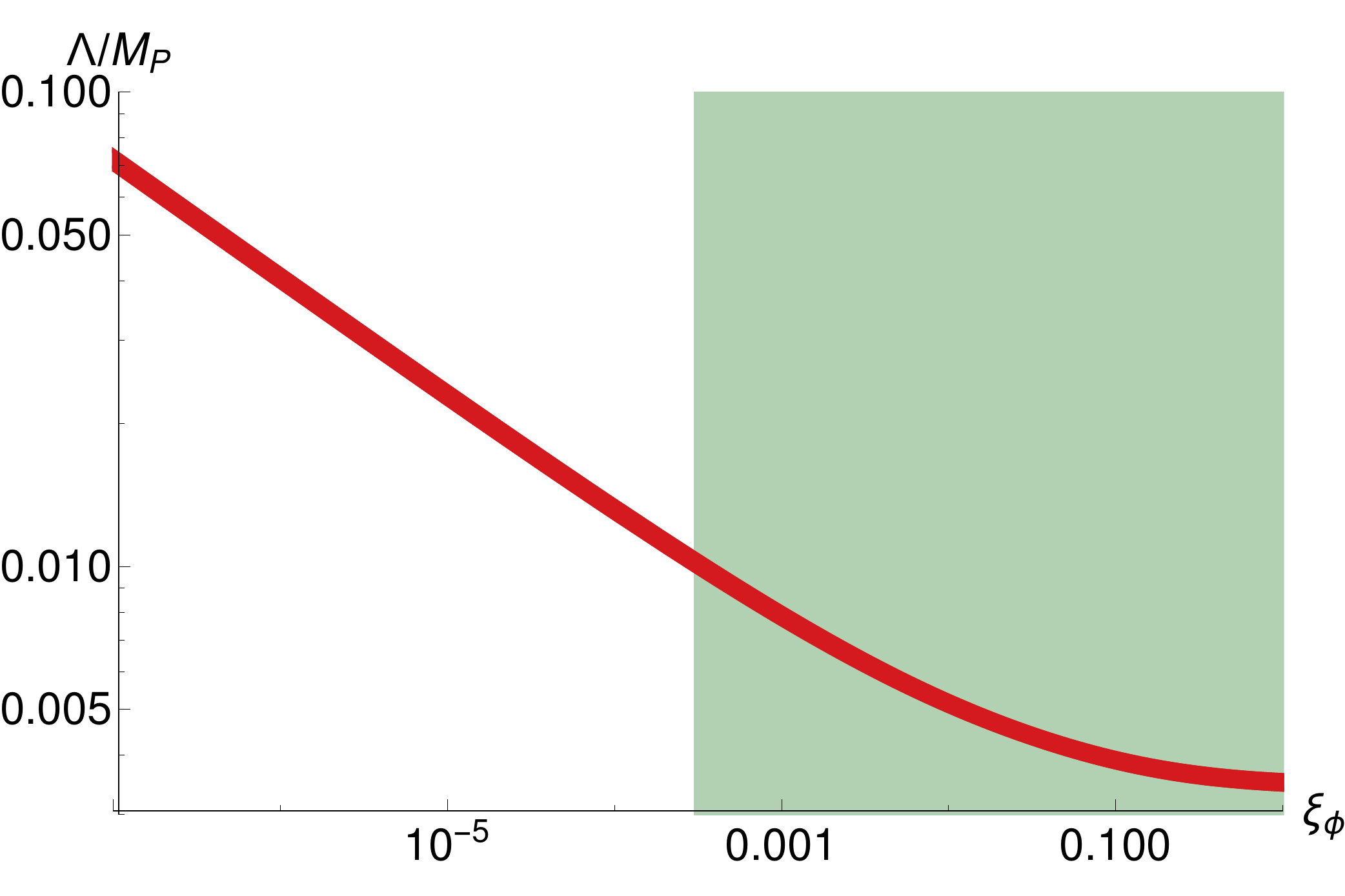}
 \quad
 \includegraphics[width=0.45\textwidth]{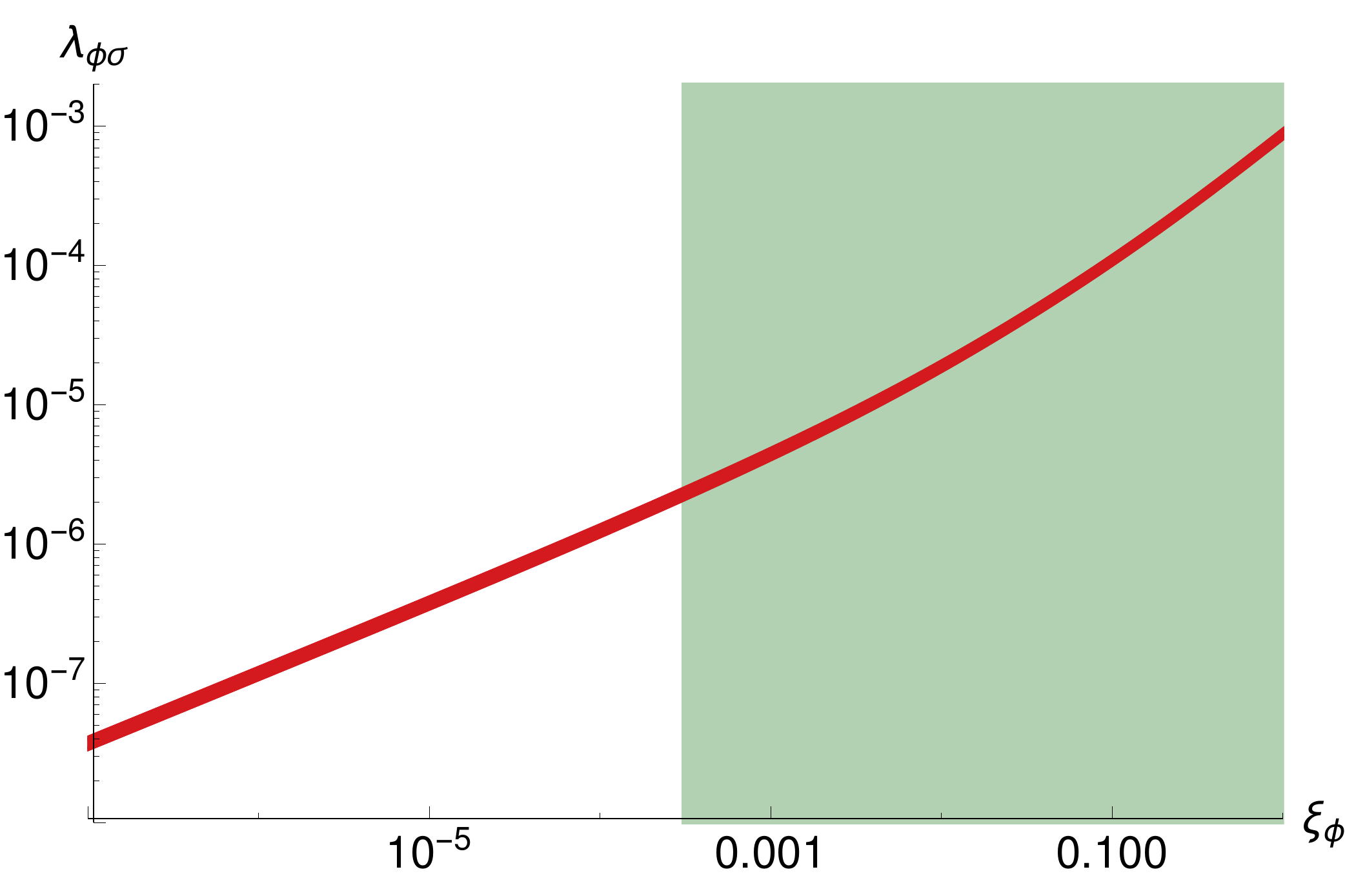}
 \caption{The values of $\Lambda$ (left) and $\lambda_{\phi\sigma}$ (right) as functions of $\xi_\phi$ in case of positive-field inflation. The light-green area is favoured by
 the present cosmological data \cite{Ade:2015tva,Ade:2015fwj,Ade:2015xua,Ade:2015lrj} .}
 \label{Fig:Lambda:vs:xi}
\end{figure}

$\Lambda$ is fixed in order to agree with the constraint on the amplitude of scalar perturbations $A_s=(\PRexp)\times 10^{-9}$. The results for $\Lambda$ as a function of $\xi_\phi$ are presented in the left panel of Fig.~\ref{Fig:Lambda:vs:xi}. From now on we will plot only results for positive-field inflation, since it is the only one in agreement with the experimental data. As expected, $\Lambda$ is always sub-Planckian.

\begin{figure}[t!]
\centering
 \includegraphics[width=0.45\textwidth]{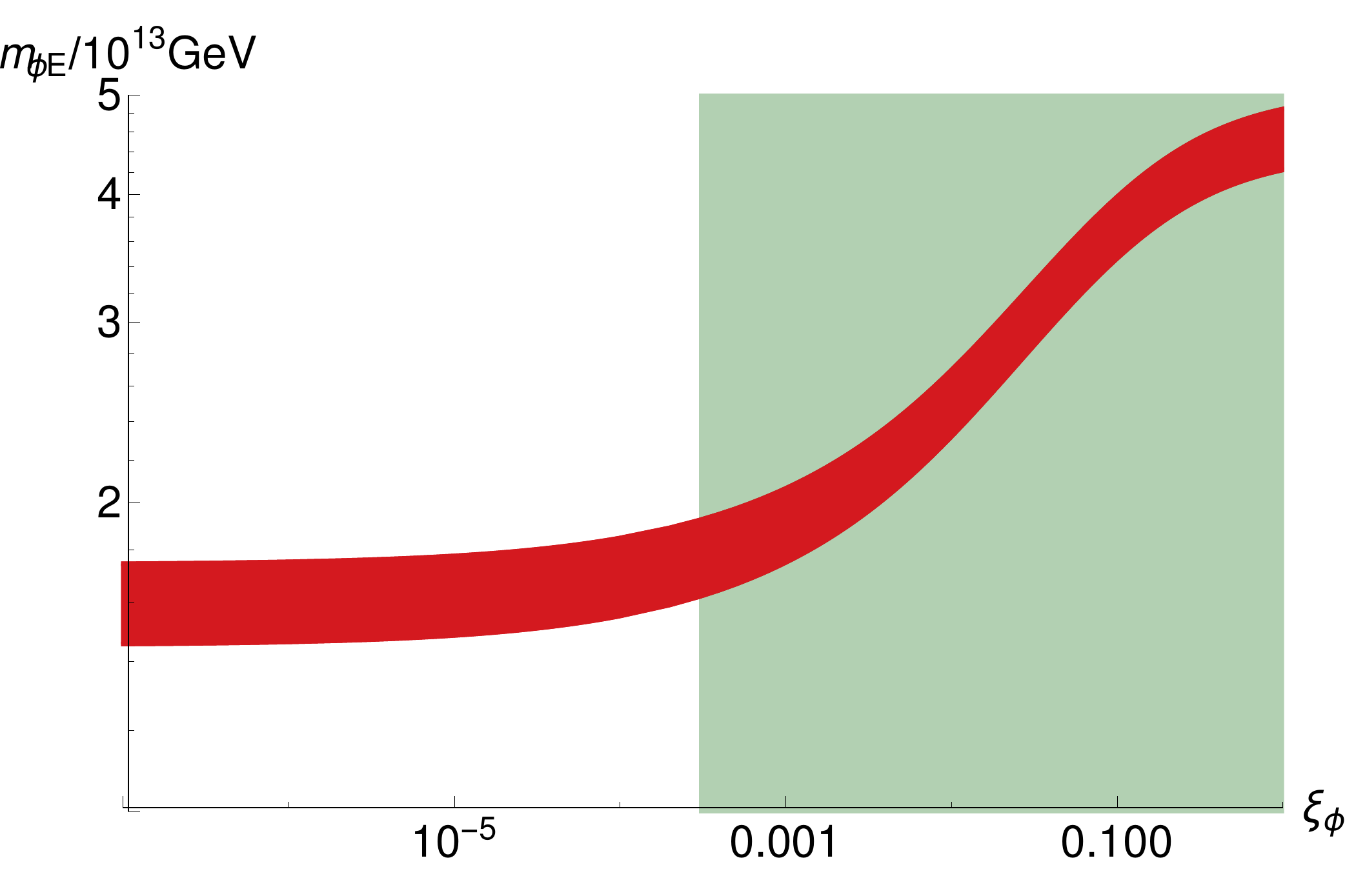}
 \quad
 \includegraphics[width=0.45\textwidth]{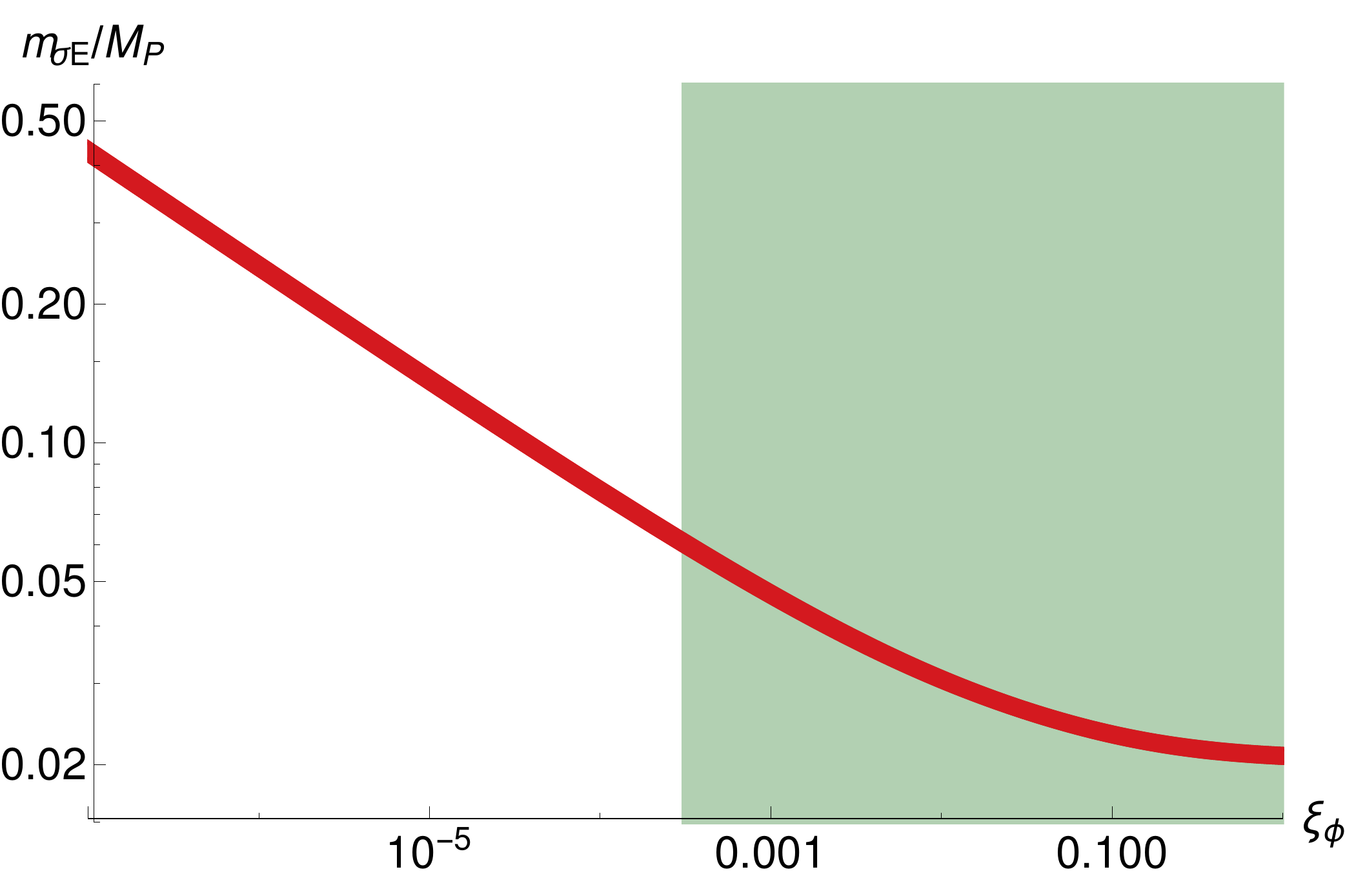}
 \caption{The values of $m_{\phi_E}$ (left) and $m_{\sigma_E}$ (right) as functions of $\xi_\phi$ in case of positive-field inflation. The light-green area is favoured by
 the present cosmological data \cite{Ade:2015tva,Ade:2015fwj,Ade:2015xua,Ade:2015lrj} .}
 \label{Fig:masses:vs:xi}
\end{figure}

To complete the model independent analyses, we present  in Fig.~\ref{Fig:masses:vs:xi} the inflaton mass $m_{\phi_E}$ (left panel) as a function of $\xi_\phi$. As usual the inflaton mass is around $10^{13}$ GeV.

\section{The minimal model of linear inflation} \label{sec:mod:dip}

The minimal model of linear inflation requires just one scalar field $\sigma$ in addition to the inflaton $\phi.$ Therefore the minimal extra-matter Lagrangian is
\begin{equation}
\mathcal{L}^{J}(\sigma,\psi,A_\mu)= \mathcal{L}^{J}(\sigma) =  \frac{(\partial \sigma)^{2}}{2} - \frac{1}{4} \lambda_{\phi \sigma} \phi^{2} \sigma^{2} - \frac{1}{4} \lambda_{\sigma} \sigma^{4} .
\end{equation}
Notice that no explicit mass term is introduced to the model. Thus all scales are generated dynamically, as we shall demonstrate in the following.
In order to compute the 1-loop potential of the inflaton we must solve the renormalisation group equations (RGEs) of the scalar couplings.
The RGEs in the Jordan frame are
\begin{align}
  16 \pi^{2} \beta_{\lambda_{\phi}} &= 18 \lambda_{\phi}^{2} + \frac{1}{2} \lambda_{\phi\sigma}^{2},
  \label{eq:blambda}
  \\
  16 \pi^{2} \beta_{\lambda_{\sigma}} &= 18 \lambda_{\sigma}^{2} + \frac{1}{2} \lambda_{\phi\sigma}^{2},
  \\
  16 \pi^{2} \beta_{\lambda_{\phi\sigma}} &= 4 \lambda_{\phi\sigma}^{2}
  + 6 \lambda_{\phi\sigma} (\lambda_{\phi} + \lambda_{\sigma}),
  \\
  16 \pi^{2} \beta_{\xi_{\phi}} &=  6 \lambda_{\phi} \left( \xi_{\phi} + \frac{1}{6} \right) + \frac{1}{3}\lambda_{\phi\sigma},
    \label{eq:beta:xiS}
    \\
\beta_{\Lambda} &= 0.
\end{align}
Because we work in the weak gravity limit~\cite{Kannike:2015apa} and the theory is massless at tree-level, at 1-loop $\Lambda$ does not receive any quantum corrections.
Because $\lambda_\phi$ runs from positive to negative values at a certain scale $\mu_0$, it is natural to consider $\lambda_\phi$ to be a subdominant coupling, so that
\begin{equation}
\lambda_{\phi} \ll \lambda_{\phi \sigma},  \quad \beta_{\xi_{\phi}} \ll \xi_{\phi}.
\end{equation}
Adding one more reasonable extra condition on $\lambda_\sigma$ we get the limit
\begin{equation}
\lambda_{\phi} \ll \lambda_{\phi \sigma} \ll \lambda_{\sigma},  \quad \beta_{\xi_{\phi}} \ll \xi_{\phi},
\label{eq:good:limit}
\end{equation}
under which the RGEs have the following simple form
\begin{align}
  16 \pi^{2} \beta_{\lambda_{\phi}} &\simeq \frac{1}{2} \lambda_{\phi\sigma}^{2},
  \label{eq:blambda:app}
  \\
  16 \pi^{2} \beta_{\lambda_{\sigma}} &\simeq 18 \lambda_{\sigma}^{2} ,
  \\
  16 \pi^{2} \beta_{\lambda_{\phi\sigma}} &\simeq  6 \lambda_{\phi\sigma} \lambda_{\sigma},
  \\
  \beta_{\xi_{\phi}} &\simeq 0,
    \label{eq:beta:xiS:app}
    \\
\beta_{\Lambda} &= 0,
\end{align}
and can be solved exactly:
\begin{eqnarray}
 \lambda_\phi(\mu)  &=& \frac{\lambda _{\phi \sigma }\left(v_{\phi}\right){}^2}{12 \lambda _{\sigma }\left(v_{\phi }\right)}
\left\{1-\frac{3 \lambda _{\sigma }\left(v_{\phi }\right)}{32 \pi ^2}-
 \left[1 -\frac{9 \lambda _{\sigma}\left(v_{\phi }\right)}{8 \pi ^2 } \log \frac{\mu }{v_\phi}\right]^{1/3}\right\} ,\qquad
 \label{eq:lambda:RGE}
\\
 \lambda_{\phi \sigma} (\mu)   &=& \lambda _{\phi\sigma }\left(v_{\phi }\right)
 \left[1 -\frac{9 \lambda _{\sigma}\left(v_{\phi }\right)}{8 \pi ^2 } \log \frac{\mu }{v_\phi} \right]^{-1/3},
\\
 \lambda_\sigma (\mu)   &=& \frac{\lambda _{\sigma }\left(v_{\phi }\right)}{1
 -\frac{9 \lambda _{\sigma}\left(v_{\phi }\right)}{8 \pi ^2 } \log \frac{\mu }{v_\phi}}.
\end{eqnarray}
We restrict ourselves to the parameter space %\footnote{It has been shown in \cite{Kannike:2015apa} that under the limit (\ref{eq:good:limit}) the RGEs for the scalar couplings in the Jordan and Einstein frame are equivalent.}
determined by (\ref{eq:good:limit}) and consider two limiting configurations: the one where the contribution of $\lambda_\sigma$ to the RGEs (and therefore to the inflationary physics) is negligible, i.e., $\lambda_\sigma (v_\phi) \ll 1$, and the one where  $\lambda_\sigma (v_\phi) \sim 1$ is close to the perturbative upper limit.

\subsection{The first configuration: $\lambda_\sigma (v_\phi) \ll 1$} \label{subsec:1stcase}
Such a configuration implies $\beta_{\lambda_{\phi\sigma}} \simeq 0 $ and, therefore, $\beta_{\lambda_{\phi}} $ is approximately constant. So the results of this configuration agree with the model independent results obtained before.
On the right panel of Fig.~\ref{Fig:Lambda:vs:xi} we plot values of the portal coupling $\lambda_{\phi\sigma}$ as a function of $\xi_\phi$. Notice that the portal remains always perturbative and much smaller than one in the region of interest, making our computations self-consistent.
To conclude, we present  in Fig.~\ref{Fig:masses:vs:xi} (right panel) the values of $\sigma_E$ mass as a function of $\xi_\phi$. As usual the inflaton mass is around $10^{13}$ GeV while the $m_{\sigma_E}$ is always much bigger than $m_{\phi_E}$ but sub-Planckian, confirming our assumption that $\sigma$ is frozen during inflation. Moreover we notice that, \emph{surprisingly,}  $m_{\sigma_E}$ has a behaviour similar to $\Lambda$ instead of $\lambda_{\phi\sigma}$. This can be easily explained by using eq.~(\ref{eq:portal}) and performing the following computation:
\begin{equation}
 m_{\sigma_E}^2 = \frac{1}{2} \lambda_{\phi \sigma} v_\phi^2 = 8 \sqrt{2} \pi  \Lambda^2 ,
\end{equation}
where we see that $m_{\sigma_E}$ just depends linearly on $\Lambda$.

\begin{figure}[t!]
\centering
 \includegraphics[width=0.7\textwidth]{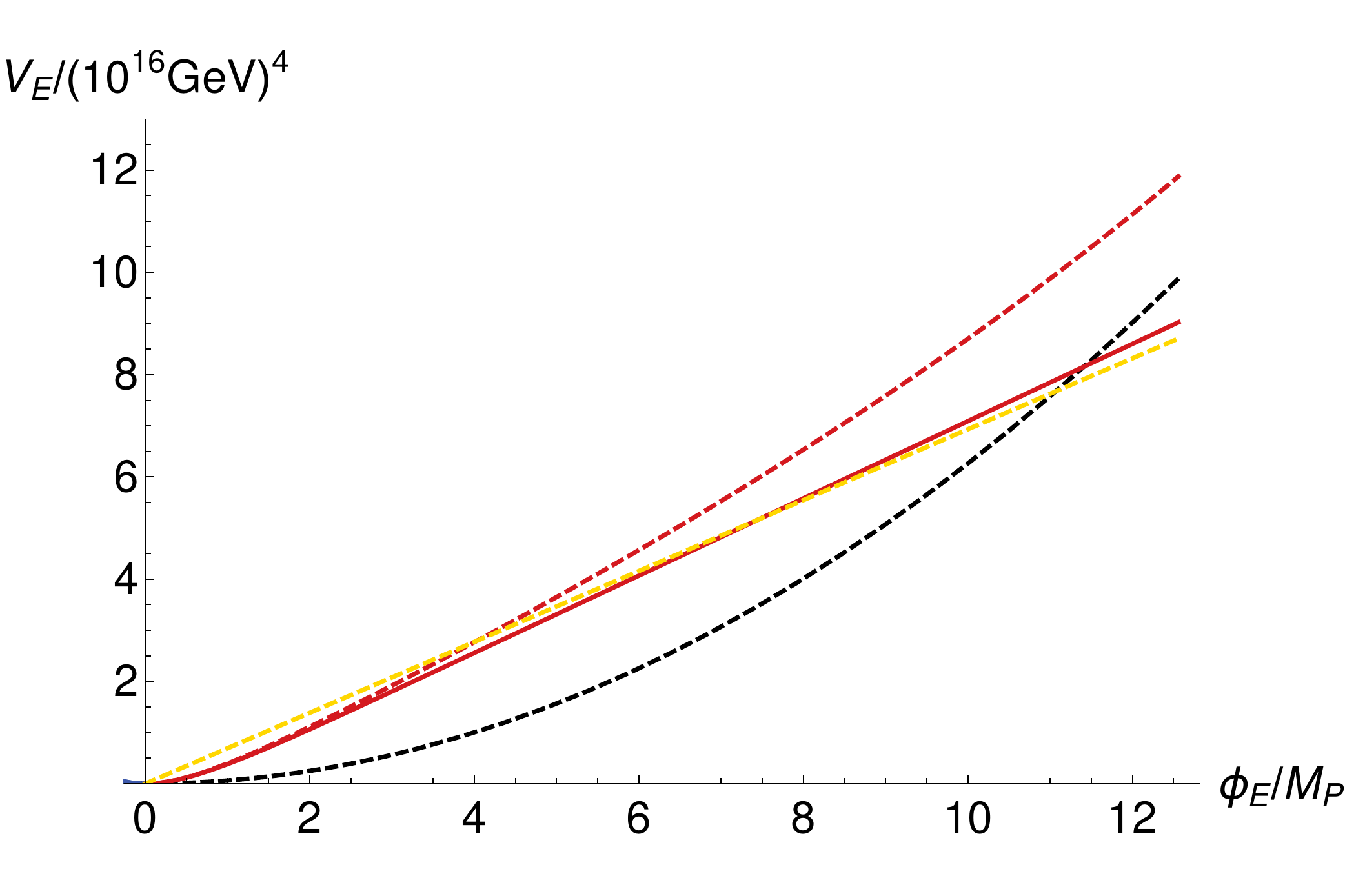}
 \caption{The potential $V_E(\phi_E)$ for $\lambda_{\sigma}(v_\phi) \simeq 1$, $\lambda_{\phi\sigma}(v_\phi) \simeq 0.016$  and $\xi_\phi\simeq 20$. The dashed red line
 corresponds to eq.~(\ref{eq:Veff:Einstein:l}) for $\phi_E>0$. The continuous lines present the scalar potential given by eq.~(\ref{eq:Veff:Einstein}) for the same parameters. The quadratic and linear  limits are presented for comparison as before.}
 \label{Vfig_l}
\end{figure}

\subsection{The second configuration: $\lambda_\sigma (v_\phi) \sim 1$} \label{subsec:2ndcase}
In this case $\lambda_\sigma (v_\phi)$ is not negligible anymore. Therefore we must work with the full expression of the running coupling $\lambda_\phi (\mu)$,  eq.~(\ref{eq:lambda:RGE}). As usual, we resum log-enhanced quantum corrections by  identifying the renormalisation scale $\mu$ with the inflaton field value $\phi$. Performing computations analogous to the one in the previous subsections, we get the following Einstein frame potential
\begin{eqnarray}
&& \hspace{-0.3cm}
V_E(\phi_E) = \label{eq:Veff:Einstein:l} \\
&& \hspace{-0.3cm}
 \frac{\lambda_{\phi \sigma }\left(v_{\phi }\right){}^2 M_P^4 }{512 \pi^2 \xi_\phi ^2}
   \left\{\frac{32 \pi ^2}{3\lambda _{\sigma }\left(v_{\phi }\right)}
   \left[1-\Bigg(1 -\frac{9 \lambda _{\sigma}\left(v_{\phi }\right)}{8 \pi ^2 } \sqrt{\frac{\xi_\phi}{1+6 \xi_\phi}} \frac{\phi_E}{M_P}\Bigg)^{1/3}\right]
         +e^{-4 \sqrt{\frac{\xi_\phi}{1+6 \xi_\phi}} \frac{\phi_E}{M_P}}-1\right\}. \nn
\end{eqnarray}
With respect to the previous case, now there is one more free parameter: $\lambda _{\sigma }\left(v_{\phi }\right)$.  We plot in Fig.~\ref{Vfig_l} the new potential for a choice of fixed free parameters, $\lambda_{\sigma}(v_\phi) \simeq 1$, $\lambda_{\phi\sigma}(v_\phi) \simeq 0.016$  and $\xi_\phi\simeq 20$. The dashed red contour corresponds to $\phi_E >0$ where the linear limit dominates. For comparison, the continuous line presents the scalar potential given by eq.~(\ref{eq:Veff:Einstein}) for the same parameters. The black dashed line presents the quadratic limit and the yellow dashed the linear limit, as before. For positive field values we can appreciate the difference between the scalar potential (\ref{eq:Veff:Einstein:l}) and the potential (\ref{eq:Veff:Einstein}), which comes from the extra free parameter $\lambda _{\sigma }$. We are not plotting the region $\phi_E<0$ becomes is essentially undistinguishable from the $\lambda_\sigma (v_\phi) \ll 1$ case, and therefore, ruled out by data ~\cite{Ade:2015tva,Ade:2015fwj,Ade:2015xua,Ade:2015lrj}.

\begin{figure}[t!]
\centering
 \includegraphics[width=0.45\textwidth]{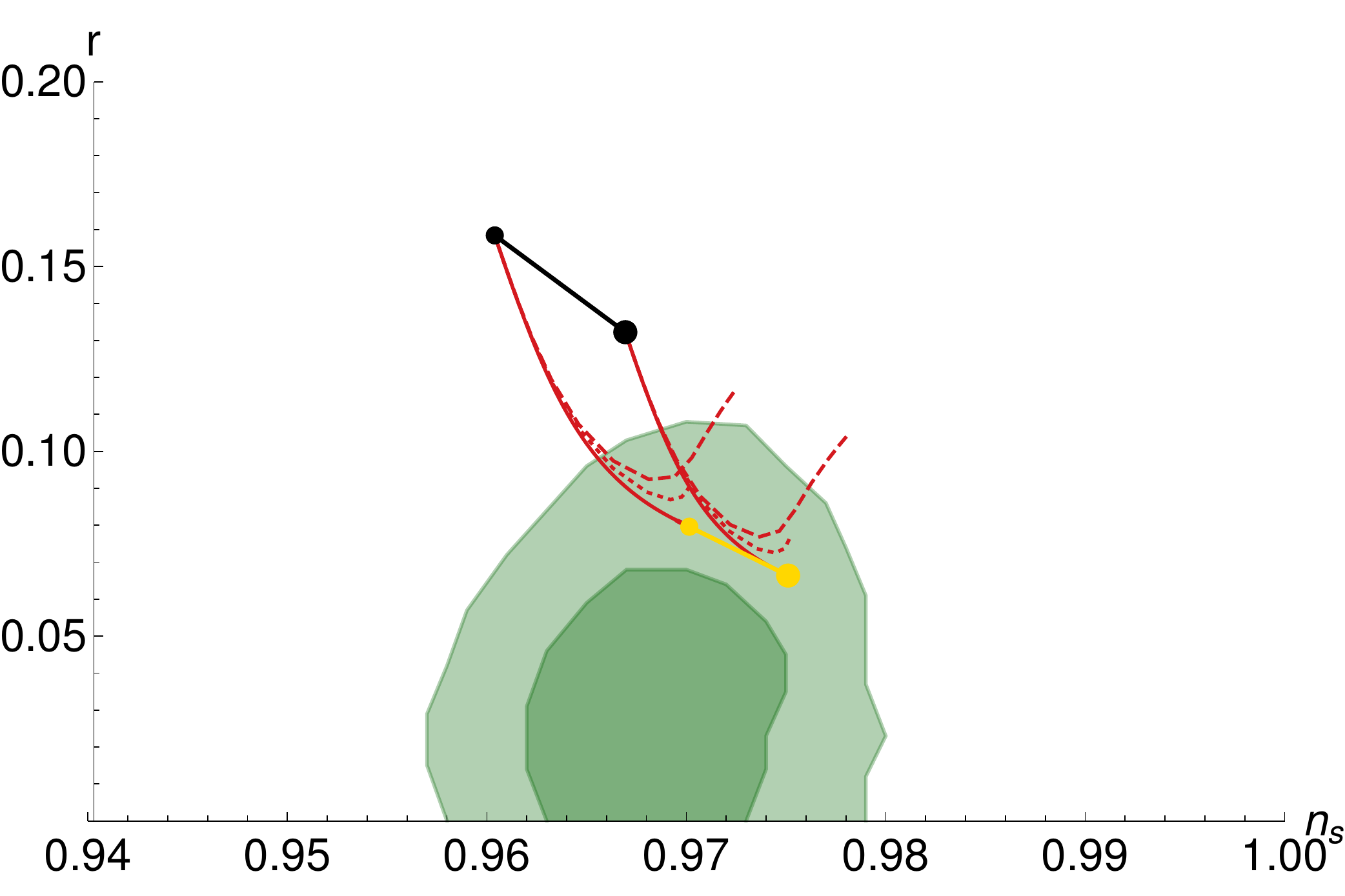}
 \quad
 \includegraphics[width=0.45\textwidth]{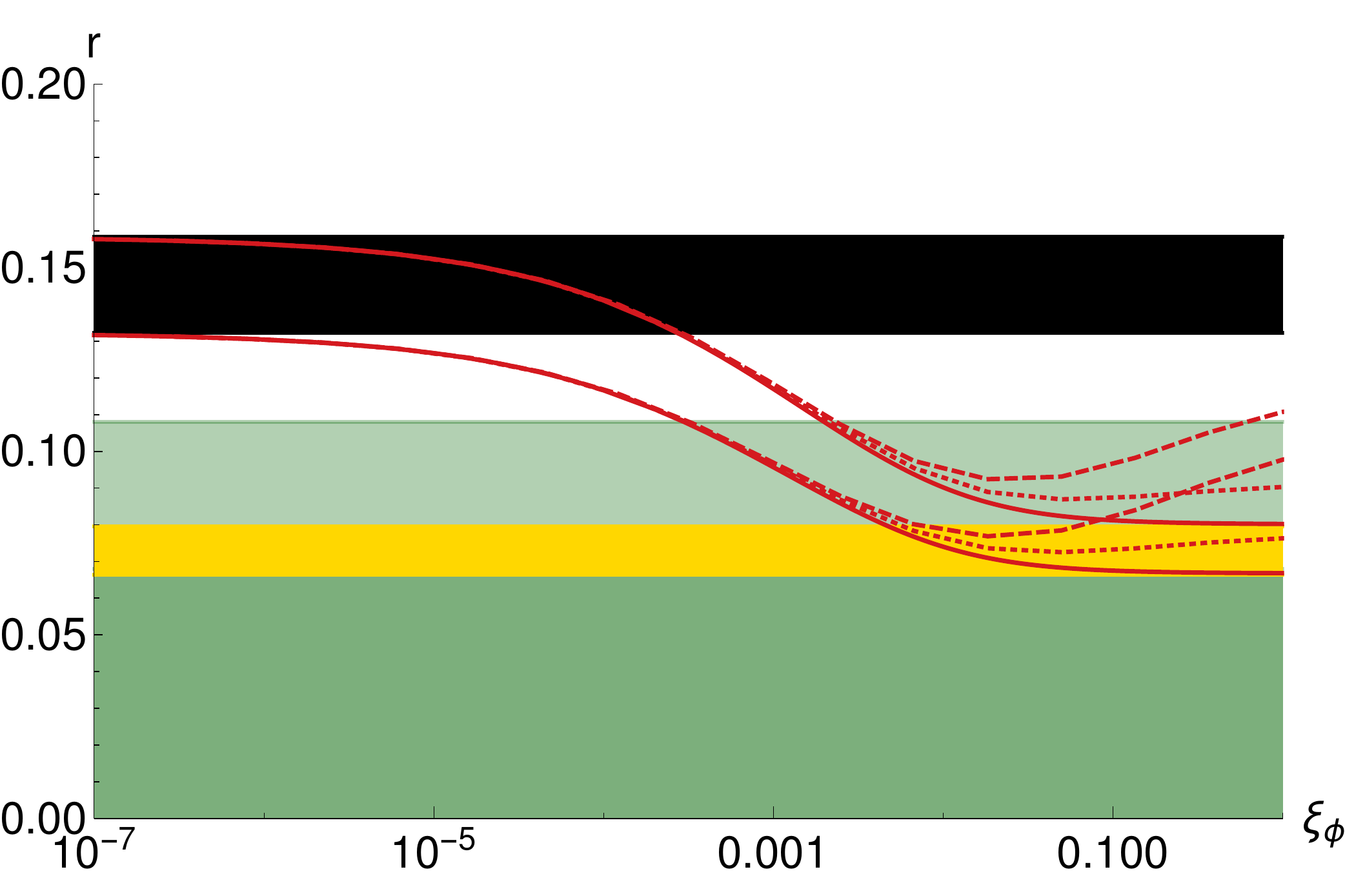}
 \caption{Predictions for tensor-to-scalar ratio $r$ for $N=50,60$ $e$-folds as a function of $n_s$ (left panel) and as a function of $\xi_\phi$ (right panel). The colour code is the same as before. The continuous line represents the limit $\lambda_{\sigma}(v_\phi) \to 0$, the dotted line $\lambda_{\sigma}(v_\phi) =0.5$ and the dashed line $\lambda_{\sigma}(v_\phi) =1$.
}
 \label{Fig:rvsn_l}
\end{figure}

To complete the analysis, we present in Fig.~\ref{Fig:rvsn_l} the predictions for tensor-to-scalar ratio $r$ in this scenario for $N=50,60$ $e$-folds as a function of $n_s$ (left panel) and as a function of $\xi_\phi$ (right panel). The colour code is the same as before. The continuous line represents the limit $\lambda_{\sigma}(v_\phi) \to 0$, the dotted line $\lambda_{\sigma}(v_\phi) =0.5$ and the dashed line $\lambda_{\sigma}(v_\phi) =1$. We neglect the predictions for negative field values because we know from the previous section that they are ruled out.
For small values of $\xi_\phi$ the predictions are the same as in the previous case, while for larger values of $\xi_\phi$ we can appreciate a difference between the two scenarios. The difference increases with increasing $\lambda_{\sigma} (v_\phi)$, and the results depart more and more from the linear limit.\footnote{{We stress that these results are instead realisation dependent. The exact realisation of the extra sector, together with very big new couplings ($\sim O(1)$) may affect the running of $\lambda_\phi$ so that the approximation in eq. (\ref{eq:lambda:phi:approx}) does not hold anymore. The exact behaviour of the RGEs will be certainly model dependent and worth investigating, but above the minimality purposes of this work. Therefore we do not consider in here any further realisation than the minimal one.}} We considered as a numerical upper bound $\xi_\phi \lesssim 400$, which keeps all the couplings perturbative at least in the region where inflation happens. Concerning the other physical parameters like the particle masses $m_\phi$, $m_\sigma$ and  $\Lambda$, the results are not so much different from the ones of the previous section, and we omit the corresponding plots.

%%%%%%%%%%%%%%%%%%%
\section{Reheating} \label{sec:Reheating}
%%%%%%%%%%%%%%%%%%%%%

Working in the Jordan frame, it may seem that reheating of the Universe in this scenario is trivial. There seem to exist inflaton couplings to extra matter fields,
 providing decays channels for reheating. However, this conclusion would be premature. Going to the Einstein frame, in which
the reheating temperature of the Universe can be computed by
\begin{equation}
  T_{\rm RH} = \left( \frac{90}{g_{*} \pi^{2}} \right)^{\frac{1}{4}} \sqrt{\Gamma_{\phi_E} M_{\rm P}},
\end{equation}
where $g_{*} \sim 100$ is the number of relativistic degrees of freedom and $\Gamma_{\phi_E}$ is the total decay width of the inflaton,
the scalar portal couplings become mass terms, and there seem to be no inflaton decay channels \cite{Kannike:2015apa}.
In addition, the inflaton decays into $\sigma$ scalars are kinematically forbidden since the latter are much heavier than the inflaton.

Fortunately,  the inflaton can decay via other channels thanks to the non-minimal coupling to gravity,
for example into the Higgs boson pairs.  The full theory Lagrangian contains, of course, also the SM terms,\footnote{Electroweak symmetry breaking may be induced in a classically scale invariant theory as, for  instance, in \cite{Gabrielli:2013hma}. However we do not specify such a mechanism here since it is beyond the aim of this work.} plus eventual BSM physics. After moving from the Jordan frame to the Einstein frame, the Higgs kinetic term (ignoring now for simplicity the gauge term of the covariant derivative) will be
\begin{equation}
\begin{split}
 S_{\text{kin},h} &=\int d^4 x \sqrt{-g^E} \frac{\partial_\mu \left( h_E \Omega \right) \partial^\mu \left( h_E \Omega \right)}{2\Omega^2}
 \\
 &\simeq  \int d^4 x \sqrt{-g^E} \Big( \frac{1}{2} \partial_\mu h_E \partial^\mu h_E +
           2 \sqrt{\frac{\xi_\phi}{1+6\xi_\phi}} \frac{h_E \partial_\mu {\phi_E} \partial^\mu h_E}{M_{\rm P}}+ \dots \Big),
\end{split}
\label{eq:chihhvertex}
\end{equation}
where we expanded $\Omega(\phi({\phi_E}))$ for ${\phi_E} \ll M_{\rm P}$ and kept only the leading order correction. The last term of eq. (\ref{eq:chihhvertex}) will induce an inflaton decay into a pair of Higgs scalars.\footnote{A contribution to it was computed in~\cite{Csaki:2014bua}.}
The decay width for the process is
\begin{equation}
 \Gamma_{{\phi_E} h h} = \frac{\xi_{\phi }}{64 \pi\left(1 + 6 \xi_{\phi } \right)} \frac{m_{{\phi_E} }^3}{ M_{\rm P}^2},
\end{equation}
where we neglected the Higgs mass since $m_h \ll m_{\phi_E}$.
The kinetic term of the SM gauge vectors is invariant under conformal transformation, therefore from there we cannot get a similar contribution. However the same contribution holds for the corresponding Goldstone bosons. Therefore, we can induce an inflaton decay to vectors through a decay into the Goldstone bosons getting
\begin{equation}
 \Gamma_{{\phi_E} Z Z} = \frac{1}{2} \Gamma_{{\phi_E} W W} = \Gamma_{{\phi_E} h h}.
\end{equation}
The same procedure can be extended to all the scalar particles of the theory, included the ones in the hidden sector. However in our minimal scenario the hidden sector contains only the kinematically forbidden scalar $\sigma$, therefore
\begin{equation}
 \Gamma_{{\phi_E}} = 4 \Gamma_{{\phi_E} h h} .
\end{equation}

\begin{figure}[t]
\begin{center}
\includegraphics[width=0.7\textwidth]{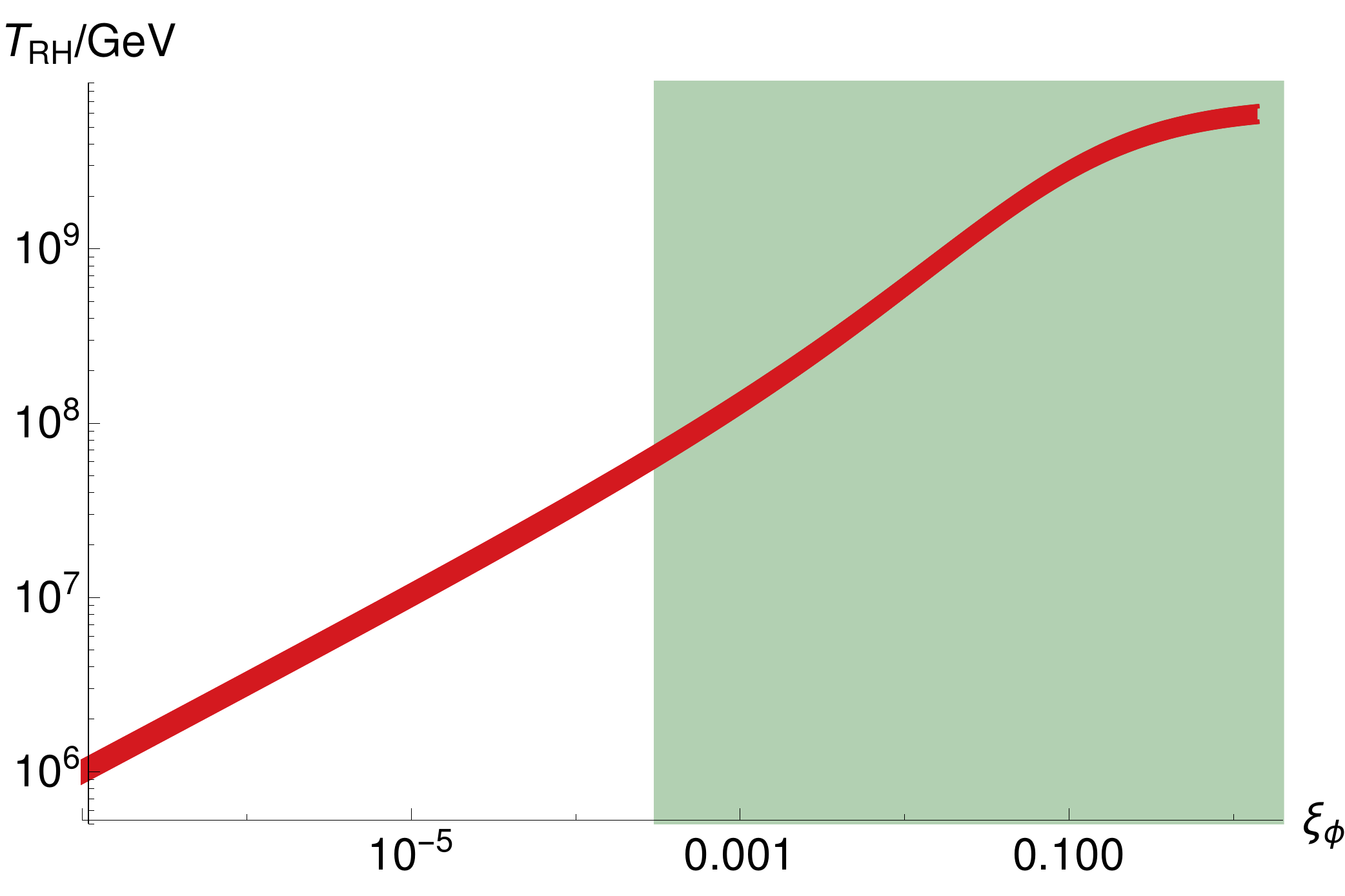}
\caption{Reheating temperature $T_{RH}$  as a function of the non-minimal coupling $\xi_\phi$, for $N=50,60$ $e$-folds, in the scenario $\lambda_\sigma (v_\phi) \ll 1$.
The reheating temperature is bounded from below by Planck data, presented by the green area.}
\label{fig:TRHvsxi}
\end{center}
\end{figure}

Another way to generate  trilinear interaction vertices that could induce the inflaton decays in our scenario is from the scaling of the explicit mass terms\footnote{Masses can be generated, again, in a classical scale invariant way adding extra scalars to the theory. For our purposes it is not needed to specify the exact mechanism, we just need to require that  such mass terms do exist. The mechanism works for all particles that get their masses from other sources than the inflaton VEV.} of the full Lagrangian (SM or other BSM physics included). For example, a fermion mass term  $m_\psi \bar \psi \psi$ in the Jordan frame  becomes in the Einstein frame
\begin{equation}
  \frac{m_\psi}{\Omega} \bar \psi_E \psi_E = \frac{m_\psi}{ e^{\sqrt{\frac{\xi_{\phi}}{1+6 \xi_{\phi}}} \frac{{\phi_E}}{M_{\rm P}}}} \bar \psi_E \psi_E \approx \left( 1 - \sqrt{\frac{\xi_{\phi}}{1 + 6 \xi_{\phi}}} \frac{{\phi_E}}{M_{\rm P}} \right) m_\psi \bar \psi_E \psi_E,
\end{equation}
where the final result is the first order approximation and $\psi_E$ is the fermion field canonically normalised in the Einstein frame.
Thus we get an induced Yukawa interaction with the coupling
\begin{equation}
 y_{{\phi_E} \bar N N} = \sqrt{\frac{\xi_{\phi}}{1 + 6 \xi_{\phi}}} \frac{m_\psi}{M_{\rm P}}.
\end{equation}
Therefore, the heavier the fermion, the bigger the coupling and the decay width.
For a decay into top quarks we would get a contribution to $\Gamma_{\phi_E}$ which is completely negligible compared to $\Gamma_{{\phi_E} h h}$. Heavy right-handed neutrinos could give a more relevant contribution. A similar computation holds also for scalars.
However for minimality we suppose that the dominant contribution to the reheating temperature comes just from the SM. In Fig. \ref{fig:TRHvsxi} we plot the reheating temperature as a  function of the non-minimal coupling $\xi_\phi$, for $N=50,60$ $e$-folds. The picture is made for the first inflation scenario for which $\lambda_\sigma (v_\phi) \ll 1$. However, the difference between the two considered scenarios will be so small to be irrelevant in the presented plot. The predicted reheating temperature range is compatible with the present constraints on reheating temperature \cite{Dai:2014jja,Munoz:2014eqa}. Notice that the Planck data, presented by the green area, imply a lower bound on the reheating temperature, which is similar to the model independent bound obtained in \cite{Domcke:2015iaa}.

%%%%%%%%%%%%%%%%%%%
\section{Discussion and Conclusions} \label{sec:Discussion}
%%%%%%%%%%%%%%%%%%%%

The recent Planck/BICEP2/Keck collaboration results~\cite{Ade:2015tva,Ade:2015fwj,Ade:2015xua,Ade:2015lrj} motivated us to generalise our previous results
on the minimal Coleman-Weinberg inflation~\cite{Kannike:2014mia} to the case of inflaton's non-minimal coupling to gravity.
In this scenario the inflaton potential as well as all particles masses are generated dynamically
via dimensional transmutation due to inflaton couplings to other fields.
 We found that (for Jordan frame large inflaton field values) the non-minimal Coleman-Weinberg scenario is confined in between two attractor solutions: quadratic inflation and linear inflation, extending the results of the previous study \cite{Rinaldi:2015yoa}.
%To our surprise, we found that the non-minimal Coleman-Weinberg scenario possesses a new attractor solution -- the linear inflation.
%To our knowledge, this result represents the first consistent 4-dimensional field-theoretic derivation of the linear inflaton potential that has %well-defined ultraviolet behaviour since it arises from the dynamics of quartic potentials.
%The previous such attempts have been based on scalar potentials that are not bounded from below or involve extra dimensions.

First we considered a model independent Coleman-Weinberg inflaton potential (\ref{eq:Veff:Jordan:Lambda}), with the inflaton non-minimally coupled to gravity and we computed which values of $r$ and $n_s$ the model can accommodate. The results show that $r$ and $n_s$ are dependent only on the assumed number of $e$-folds and the value of the non-minimal coupling $\xi_\phi$. We also noticed that the predictions are in between two attractor solutions: for $\xi_\phi$ approaching zero we get quadratic inflation values, while for
finite $\xi_\phi$  we reproduce the results of linear inflation. The latter limit is achieved for relatively small values of $\xi_\phi$: $\xi_\phi \gtrsim 0.1$.

Then we presented an explicit model for the non-minimal Coleman-Weinberg potential which involves an extra scalar $\sigma$. We computed RGEs for this model and showed for what range of the parameters we reproduce the model independent results, and for what range we depart from them. We discovered that for large values of the $\sigma$ self-quartic coupling $\lambda_\sigma$ the quadratic limit is still present but the results may depart from the linear limit.

Finally we computed the decay rates of the inflaton in this scenario and studied the reheating temperature of the Universe. We showed  that, in the minimal configuration, the reheating happens via the non-minimal coupling $\xi_\phi$ into pairs of electroweak bosons (Higgses, $Z$'s  and $W$'s) with a temperature around $10^{8-10}$ GeV.

We conclude that the non-minimal Coleman-Weinberg inflation is  attractive, self-consistent framework to address physics at the early universe.

\acknowledgments

The authors thank Alessandro Strumia and Alberto Salvio for useful discussions.
This work was supported by  the grant IUT23-6  of the Estonian Ministry of Education and Research, the Estonian Research Council grant PUT799, CERN+, and by EU through the ERDF  CoE program.

\clearpage

\appendix

\section{Computation of slow roll parameters} \label{appendix}

In this Appendix we give some more details about inflationary computations performed in Section \ref{sec:CSI_and_inflation}.
For simplicity, we perform the inflationary computations in the Einstein frame. The slow roll parameters are defined as
\begin{eqnarray}
\epsilon &=& \frac{M_P^2}{2} \left( \frac{V_E'}{V_E} \right)^2,\\
\eta &=& M_P^2 \frac{V_E''}{V_E}.
\end{eqnarray}
where, according to the configuration of interest, $V_E$ is the inflaton potential given by eq. (\ref{eq:Veff:Einstein}) or (\ref{eq:Veff:Einstein:l}), and the symbol `` $'$ '' represents the derivative with respect to $\phi_E$.
From now on, we split the discussion according to the two different configurations described respectively in Section \ref{subsec:1stcase} and Section \ref{subsec:2ndcase} (i.e. potential (\ref{eq:Veff:Einstein}) or (\ref{eq:Veff:Einstein:l}) ).

\subsection{The first configuration: $\lambda_\sigma (v_\phi) \ll 1$}
The slow roll parameters are
\begin{eqnarray}
\epsilon &=& \frac{8 \xi_\phi }{\left(1 + 6 \xi_\phi \right)}
\left(\frac{e^{-4\sqrt{\frac{\xi_{\phi }}{1 + 6 \xi_\phi}}\frac{\phi_E}{M_P}}-1}{
4\sqrt{\frac{\xi_{\phi }}{1 + 6 \xi_\phi}}\frac{\phi_E}{M_P}+e^{-4\sqrt{\frac{\xi_{\phi }}{1 + 6 \xi_\phi}}\frac{\phi_E}{M_P}}-1} \right)^2, \\
&&\quad\nonumber \\
&&\quad\nonumber \\
\eta &=& \frac{16  \xi_\phi }{\left(1 + 6 \xi_\phi\right)}
\frac{1}{1+e^{4 \sqrt{\frac{ \xi_\phi }{1 + 6 \xi_\phi}} \frac{\phi_E }{M_P}} \left(4 \sqrt{\frac{ \xi_\phi }{1 + 6 \xi_\phi}} \frac{\phi_E }{M_P}-1\right)}.
\end{eqnarray}

\subsection{The second configuration: $\lambda_\sigma (v_\phi) \sim 1$}
The slow roll parameters are
\begin{eqnarray}
\epsilon &=&  \frac{8 \xi_\phi}{1+6 \xi_\phi}
\left\{ \frac{e^{-4 \sqrt{\frac{\xi_\phi }{1 + 6 \xi_\phi}} \frac{\phi_E}{M_P}}-
   \left[1-\frac{9  \lambda _{\sigma }\left(v_{\phi }\right)}{8 \pi ^2} \sqrt{\frac{\xi_\phi }{1 + 6 \xi_\phi}} \frac{\phi_E}{M_P}\right]^{-\frac{2}{3}}}{
   1-e^{-4 \sqrt{\frac{\xi_\phi }{1 + 6 \xi_\phi}} \frac{\phi_E}{M_P}}+\frac{32 \pi ^2}{3 \lambda _{\sigma }\left(v_{\phi
   }\right)} \left[\left(1-\frac{9 \lambda _{\sigma }\left(v_{\phi }\right)}{8 \pi ^2} \sqrt{\frac{\xi_\phi }{1 + 6 \xi_\phi}}  \frac{\phi_E}{M_P}\right)^{\frac{1}{3}}-1\right] } \right\}^2\!\!\!,\,\,\, \\
&&\quad\nonumber \\
&&\quad\nonumber \\
\eta &=& \frac{16  \xi_\phi}{\left(1 + 6 \xi_\phi\right)}
\frac{e^{-\frac{4 \phi_E \sqrt{\frac{ \xi_\phi }{1 + 6 \xi_{\phi }}}}{M_P}}+\frac{3 \lambda _{\sigma }\left(v_{\phi }\right)}{16 \pi^2}
\left[1-\frac{9 \lambda _{\sigma }\left(v_{\phi }\right)}{8 \pi ^2} \sqrt{\frac{\xi_\phi }{1 + 6 \xi_\phi}}  \frac{\phi_E}{M_P}\right]^{-\frac{5}{3}}}{e^{-\frac{4
   \phi_E \sqrt{\frac{ \xi_\phi }{1 + 6 \xi_\phi}}}{M_P}}+\frac{32 \pi ^2}{3 \lambda_{\sigma }\left(v_{\phi }\right)} \left\{1-\left[1-\frac{9 \lambda _{\sigma }\left(v_{\phi }\right)}{8 \pi ^2} \sqrt{\frac{\xi_\phi }{1 + 6 \xi_\phi}}  \frac{\phi_E}{M_P}\right]^{\frac{1}{3}}\right\} -1 }.
\end{eqnarray}

\bigskip

Then according to the configuration of interest, the tensor-to-scalar ratio $r$ and the scalar spectral index $n_s$ can be computed using the corresponding equations of the slow roll parameters:
\begin{eqnarray}
r &=& 16 \epsilon, \label{eq:r}\\
n_s &=& 1 -6 \epsilon + 2 \eta \label{eq:n_s}.
\end{eqnarray}

\bibliographystyle{JHEP}
\bibliography{citations}

\end{document}